\pdfoutput=1
\documentclass[sigconf]{acmart}

\usepackage{url}
\usepackage{tikz-qtree}
\usepackage{booktabs} 
\usepackage{pgfplots}
\pgfplotsset{compat=1.17}
\usetikzlibrary{arrows.meta}
\renewcommand\footnotetextcopyrightpermission[1]{} 
\setcopyright{none}
\begin{document}
\title{A Survey of Miss Ratio Curve Construction Techniques}

\author{Daniel Byrne}
\affiliation{%
  \institution{Michigan Technological University}
  \city{Houghton}
  \state{Michigan}
  \country{USA}
}
\email{djbyrne@mtu.edu}

\renewcommand{\shortauthors}{Byrne D.}

\begin{abstract}
Miss ratio curve (MRC), or equivalently hit ratio curve (HRC), construction
techniques have gathered the attention of many researchers. Advances in
sampling now allow these curves to be approximated in constant space,
which makes online working-set-size (WSS) measurement practical. Techniques span the
algorithmic design paradigm from classic stack processing to
compiler analysis and machine learning. Our survey produces a broad classification
of the techniques based primarily on \emph{what} locality
metric is recorded and \emph{how} that metric is stored
for processing.

Applications of these curves span from dynamic cache partitioning
in the processor, to slab allocation in key-value caches, to memory tiering
at the operating system level. Our survey gives an overview of
the historical, exact MRC construction methods, and compares them
with the state-of-the-art methods present in today's literature.
This revised version improves our coverage of existing techniques and adds the techniques
published since 2017 (including methods for non-LRU and multi-tier caches), the unification of the
reuse time models, and a short introduction to writeback ratio curves.
Finally, we revisit the open problems posed in the first version and
show where there are still open areas of research.
\end{abstract}


\maketitle
\pagestyle{plain}

\section{Introduction}

From the working set theory
proposed by Denning in 1968 \cite{denning1968working} to recent advancements
in cache locality theory \cite{xiang2013hotl,hu2016kinetic,yuan2019relational},
modeling data locality has been essential to cache
design and OS resource allocation choices. In the past two decades there
has been a large emphasis on using these models to construct plots relating
miss ratio to cache size, called miss ratio curves (MRCs) or equivalently
hit ratio curves (HRCs).

Miss ratio curves have proven to be extremely useful in estimating
\emph{how much} data is being used by a particular workload, known as
the working set size. Knowing the working set size of a workload
helps estimate the \emph{utility} of adding more memory or increasing
the cache size. An example miss ratio curve is shown in Figure \ref{fig:mrc}.
Reading the curve from left to right, each point gives the fraction of requests that
would miss in an LRU cache of that size. Up to about 3,500 objects the curve is steep:
every additional 500 objects removes 7 to 15 percentage points of misses. The slope then
falls quickly, and after the \emph{knee} at about 4,500 objects the curve is nearly flat.
Growing the cache from 5,000 to 8,000 objects removes only half a percentage point, because
almost all of the remaining misses are requests to the 200,000 cold objects, which are
rarely reused soon enough to hit. Note that although only 3,000 objects are hot in this
workload, the knee lies closer to 4,500 objects because the cold objects also take up space
in an LRU cache.

\begin{figure}
\centering
\begin{tikzpicture}
\begin{axis}[
    width=\columnwidth, height=5.4cm,
    xmin=0, xmax=8000, ymin=0, ymax=1.0,
    xlabel={Cache size (objects)}, ylabel={Miss ratio},
    xtick={0,1000,2000,3000,4000,5000,6000,7000,8000},
    xticklabels={0,1k,2k,3k,4k,5k,6k,7k,8k},
    ytick={0,0.2,0.4,0.6,0.8,1.0},
    grid=major, grid style={gray!20},
    tick label style={font=\footnotesize}, label style={font=\footnotesize},
    clip=false]
  \fill[gray!12] (axis cs:4500,0) rectangle (axis cs:8000,1.0);
  \node[font=\scriptsize, align=center, text=black!70] at (axis cs:6250,0.62)
      {flat: more cache\\ small impact in miss ratio};
  \node[font=\scriptsize, align=center, text=black!70] at (axis cs:2350,0.83)
      {steep: more cache\\ large decrease in miss ratio};
  \addplot[very thick, blue!70!black] table[x=c, y=mr] {figures/data/fig1_mrc.dat};
  \draw[dashed, black!60] (axis cs:4500,0) -- (axis cs:4500,1.0);
  \fill[red!70!black] (axis cs:4500,0.1155) circle (2pt);
  \node[font=\scriptsize, anchor=south west, red!70!black, fill=gray!12, inner sep=1pt] at (axis cs:4550,0.14)
      {knee $\approx$ 4.5k objects};
  \draw[dotted, thick, black!60] (axis cs:0,0.082) -- (axis cs:8000,0.082);
  \node[font=\scriptsize, anchor=south west, black!70, fill=white, inner sep=1pt] at (axis cs:150,0.09)
      {first accesses (8.2\%)};
  \draw[{Stealth}-{Stealth}, black!70] (axis cs:5000,0.25) -- (axis cs:8000,0.25)
      node[midway, above, font=\scriptsize] {5k$\rightarrow$8k: $-$0.5 points};
\end{axis}
\end{tikzpicture}
\caption{An example miss ratio curve for an LRU cache. The workload is synthetic:
1 million requests, 90\% of which go to 3,000 hot objects (Zipf, $\alpha=0.3$) and 10\%
to 200,000 cold objects chosen uniformly. Before the knee, adding cache space keeps more
hot objects and removes many misses. After the knee, nearly all misses are requests to
cold objects, most of them being first accesses (dotted line), so additional cache capacity
barely decreases the miss ratio.}
\Description{A miss ratio curve that falls steeply from 1.0 at cache size zero to about
0.12 at 4,500 objects and then flattens, approaching the 8.2\% share of first accesses.}
\label{fig:mrc}
\end{figure}

Given the theoretic and implementation advancements in miss ratio curve
construction, we find a survey on these methods to be a useful update to the
community, and we present some identified open lines of research yet to be
fully explored in this domain.

\section{Taxonomy}

In this section we present a taxonomy of the miss ratio curve construction
techniques based on \emph{which} metric is recorded, \emph{how} the data is stored, and the
\emph{derivation method} of the miss ratio curve.

\subsection{Definitions}

In the cache modeling literature there are variations on the meaning of terms such as
\emph{reuse distance}. To avoid confusion, we give definitions for the following
terms used in our taxonomy. Throughout the survey, $N$ is the total number of
references in a trace, $M$ is the number of distinct data (objects, blocks or cache lines)
in the trace, and $c$ is the cache size measured in data items.

\begin{itemize}
    \item Stack Distance - The position of a datum in the LRU stack when it is
          referenced, counting the top of the stack as 1. Equivalently, it is the number of
          \emph{distinct} data accessed since the previous access to the same datum,
          including the datum itself. This is also referred to as reuse distance or
          LRU stack distance. A reference hits in an LRU cache of size $c$ if and only if its
          stack distance is at most $c$. The first reference to a datum has infinite stack distance.
    \item Reuse Time - The difference between the access times of two consecutive accesses
          to the same datum. It counts every access in between plus the second access itself,
          so it is never smaller than the stack distance. This is also referred to as
          reuse interval, and is consistent with \cite{hu2016kinetic,hu2015lama,zhong2009reusedist,xiangWSS}.
          We write $P(rt > t)$ for the fraction of references whose reuse time is greater than $t$.
    \item Footprint - The average number of distinct data accessed in a window
          of $x$ consecutive references, denoted $fp(x)$ \cite{xiangWSS}. It is the
          cache locality theory equivalent of Denning's working set size.
    \item Buckets - Groups of data that are next to each other in the LRU stack. If we only
          know which bucket a datum is in, we know its stack distance only to within the
          bucket's range.
    \item Access Counters - Hardware or software monitors that count unique or total
          accesses to a piece of data.
\end{itemize}

\subsection{Taxonomy of MRC Constructions}

\tikzset{edge from parent/.style=
{draw, edge from parent path={(\tikzparentnode.south) -- +(0,-8pt) -| (\tikzchildnode)}},
blank/.style={draw=none}}

Figures \ref{fig:rd} and \ref{fig:rt} classify the
techniques that use stack distance and reuse time respectively,
by how they store and process the data into a miss ratio curve. Each figure is organized
from top to bottom by metric (stack distance or reuse time), primary data structure or
mechanism, miss ratio curve derivation method, and representative works.
Since the first version of this survey, a third class of techniques has emerged
that records neither metric directly, but instead simulates or models the cache state of
an arbitrary replacement policy. We classify these techniques in Figure \ref{fig:any}
and describe them in Section \ref{sec:nonlru}. Table \ref{tab:glance} summarizes
representative techniques from all three classes, including the kind of accuracy
guarantee each one offers.

\begin{table}[t]
\footnotesize
\centering
\caption{Representative techniques at a glance. SD = stack distance, RT = reuse time,
IRM = independent reference model. Space excludes the trace itself.}
\label{tab:glance}
\setlength{\tabcolsep}{4pt}
\begin{tabular}{@{}llll@{}}
\toprule
\textbf{Technique} & \textbf{Metric} & \textbf{Space} & \textbf{Accuracy guarantee} \\
\midrule
Mattson \cite{mattson1970evaluation}      & SD & $O(M)$ & exact \\
Olken \cite{olken1981efficient}           & SD & $O(M)$ & exact \\
Scale tree \cite{ding2003predicting}      & SD & $O(\log M)$ tree$^{*}$ & deterministic, relative $e$ \\
SHARDS \cite{waldspurger2015efficient}    & SD & $O(1)$ & statistical \\
Counter Stacks \cite{wires2014characterizing} & SD & sublinear & probabilistic counters \\
Drudi \cite{drudi2015approximating}       & SD & sublinear & randomized, additive $\epsilon$ \\
Incr.-and-Freeze \cite{bender2023iaf}     & SD & $O(N)$ & exact \\
HayStack \cite{gysi2019haystack}          & SD & program size & exact for affine loops \\
\midrule
StatCache \cite{berg2004statcache}        & RT & $O(1)$, sampled & model (random repl.) \\
StatStack \cite{eklov2010statstack}       & RT & $O(1)$, sampled & conditional (\S\ref{sec:unify}) \\
Footprint \cite{xiangWSS}                 & RT & $O(M)$ & conditional (\S\ref{sec:unify}) \\
AET \cite{hu2016kinetic}                  & RT & $O(1)$, sampled & conditional (\S\ref{sec:unify}) \\
Write footprint \cite{smith2021writeback} & RT & none (Zipf) & model (IRM) \\
\midrule
MiniSim \cite{waldspurger2017minisim}     & any policy & $O(1)$ per size & empirical \\
Kosmo \cite{shakiba2024kosmo}             & 6 policies & sampled & empirical \\
TTL approx.\ \cite{che2002hierarchical}   & analytical & none & asymptotic (IRM) \\
\bottomrule
\end{tabular}

\vspace{2pt}
{\footnotesize $^{*}$ plus the $O(M)$ table of last access times needed by every tree method.}
\end{table}

\begin{figure*}[t]
\centering
{%
\begin{tikzpicture}[every tree node/.style={align=center, font=\scriptsize},
                    level distance=0.95cm, sibling distance=1pt]
\Tree
    [.{\textbf{Stack Distance}}
        [.{Tree}
            [.{Exact\\Tree}
                {Bennett \cite{bennett1975lru}\\Almási \cite{Almasi}\\Olken \cite{olken1981efficient}} ]
            [.{Approx.\\Tree}
                {\textbf{Scale Tree}\\\cite{ding2003predicting,zhong2009reusedist}} ]
            [.{Sampled\\Tree}
                {SHARDS \cite{waldspurger2015efficient}\\Carra \cite{carra2020efficient}\\FLOWS \cite{guo2024flows}} ] ]
        [.{Parallel}
            [.{Chunks /\\Divide \& Conquer}
                {PARDA \cite{parda}\\HP-RDA \cite{cui2012hprda}\\Incr.-and-Freeze \cite{bender2023iaf}} ] ]
        [.{Buckets}
            [.{Stack Dist.\\Histogram}
                {MIMIR \cite{saemundsson2014cloudcaches}} ]
            [.{Hill\\Climbing}
                {Cliffhanger \cite{cidon2016cliffhanger}\\Memshare \cite{cidon2017memshare}} ] ]
        [.{Histogram}
            [.{Stack\\Processing}
                {Kim \cite{Kim}\\UMONs \cite{qureshi2006utility}} ]
            [.{Cardinality\\Sketch}
                {Counter Stacks \cite{wires2014characterizing}\\CounterStacks++ \cite{sultan2024ttls}\\Drudi \cite{drudi2015approximating}} ]
            [.{Inter-miss\\Gap}
                {FractalMRC \cite{fractal}} ] ]
        [.{Symbolic}
            [.{Polyhedral\\Counting}
                {PolyCache \cite{bao2018polycache}\\HayStack \cite{gysi2019haystack}} ] ]
    ]
\end{tikzpicture}}
\caption{Methods based on stack distance. From top to bottom: metric, primary data
structure or mechanism, MRC derivation method, and representative works.}
\Description{A tree diagram classifying stack distance methods into tree, parallel,
bucket, histogram and symbolic methods, with representative works at the leaves.}
\label{fig:rd}
\end{figure*}

\begin{figure*}[t]
\centering
{%
\begin{tikzpicture}[every tree node/.style={align=center, font=\scriptsize},
                    level distance=0.95cm, sibling distance=6pt]
\Tree
    [.{\textbf{Reuse Time}}
        [.{Histogram}
            [.{Average\\Eviction}
                {AET \cite{hu2016kinetic,hu2018fast}\\pRedis \cite{pan2019predis}} ]
            [.{Footprint}
                {HOTL \cite{xiang2013hotl}\\LAMA \cite{hu2015lama}\\Write Footprint \cite{smith2021writeback}} ]
            [.{Hit-Evict\\Distribution}
                {Beyond LRU \cite{beckmann2016modeling}} ] ]
        [.{Access Counters}
            [.{Expected\\Reuse}
                {StatCache \cite{berg2004statcache}\\StatStack \cite{eklov2010statstack}\\RDX \cite{wang2019rdx}} ]
            [.{Re-access\\Ratio}
                {OSCA \cite{zhang2020osca}\\Cuki \cite{gu2023cuki}} ] ]
        [.{Static Analysis}
            [.{Static\\Sampling}
                {SPS \cite{chen2018sps}\\PLUSS \cite{liu2024pluss}} ] ]
    ]
\end{tikzpicture}}
\caption{Methods based on reuse time. AET, the footprint and StatStack are three algorithms
for the same model \cite{yuan2019relational} (Section \ref{sec:unify}).}
\Description{A tree diagram classifying reuse time methods into histogram, access counter
and static analysis methods, with representative works at the leaves.}
\label{fig:rt}
\end{figure*}

\begin{figure*}[t]
\centering
{%
\begin{tikzpicture}[every tree node/.style={align=center, font=\scriptsize},
                    level distance=0.95cm, sibling distance=6pt]
\Tree
    [.{\textbf{Cache State (any policy)}}
        [.{Per-size Simulation}
            [.{Mini-caches}
                {MiniSim \cite{waldspurger2017minisim}\\FLOWS \cite{guo2024flows}} ]
            [.{Multi-tier}
                {eMRC \cite{liu2021emrc}} ] ]
        [.{One Pass, All Sizes}
            [.{Eviction\\Maps}
                {Kosmo \cite{shakiba2024kosmo}} ]
            [.{Probabilistic\\Stack}
                {KRR \cite{yang2021krr}} ] ]
        [.{Analytical}
            [.{TTL\\Approximation}
                {Che \cite{che2002hierarchical}\\LRU(m) \cite{gast2017lrum}\\CLOCK(K) \cite{wang2026clockk}} ]
            [.{Reuse-based\\Models}
                {Sen \& Wood \cite{sen2013}} ] ]
        [.{Learned}
            [.{Learned\\Model}
                {LPCA \cite{gu2022lpca}} ] ]
    ]
\end{tikzpicture}}
\caption{Methods that build miss ratio curves for arbitrary (including non-LRU)
replacement policies, classified by how the cache state is obtained.}
\Description{A tree diagram classifying methods for arbitrary replacement policies into
per-size simulation, one-pass reconstruction, analytical and learned methods.}
\label{fig:any}
\end{figure*}

\section{Stack Distance Algorithms}

This class of algorithms takes advantage of a property of the least recently used (LRU)
policy, called the \emph{inclusion} or \emph{stack} property: at any moment, an LRU cache of
size $c$ holds exactly the $c$ most recently used data, so a smaller cache always holds a
subset of what a larger one holds. We can therefore keep one list of all data ordered by
recency, the LRU stack, and get the result for every cache size at once: a reference hits in
a cache of size $c$ if and only if its position in the stack is at most $c$.
Table \ref{tab:sd} shows the stack for a short sequence of requests.

Most methods in this survey model a fully associative cache. Hardware caches are set
associative, and a common practice is to treat each set as an independent, smaller LRU cache.
Sen and Wood \cite{sen2013} model set-associative caches with LRU, pseudo-LRU, random and
not-most-recently-used replacement directly from reuse information, and the footprint has been
extended to associativity and sector caches \cite{mappedfp2018}.

\begin{table}[t]
    \small
    \centering
    \caption{Example stack distance calculation for a set of requests. The stack
    is shown \emph{before} each request, most recent first; the stack distance
    is the position of the requested item in that stack.}
    \begin{tabular}{llll}
        \toprule
        \textbf{request} & \textbf{time} & \textbf{stack before} & \textbf{stack dist.} \\
        \midrule
        a & 1 & (empty)  & $\infty$ \\
        b & 2 & a        & $\infty$ \\
        c & 3 & b,a      & $\infty$ \\
        d & 4 & c,b,a    & $\infty$ \\
        a & 5 & d,c,b,a  & 4 \\
        a & 6 & a,d,c,b  & 1 \\
        d & 7 & a,d,c,b  & 2 \\
        b & 8 & d,a,c,b  & 4 \\
        \bottomrule
    \end{tabular}
    \label{tab:sd}
\end{table}

To generate a miss ratio curve, a \emph{histogram} of stack distances is created.
Given the stack distance histogram, one can calculate the number of hits in a cache
of size $c$ as the sum of all the references with stack distances
$\le c$, and similarly the number of misses as the sum of all references
with stack distance $> c$.
Formally, let $freq(i)$ be the number of references with stack distance $i$
(including $i=\infty$ for first references). Then the miss ratio for a cache of size $c$ is:

\begin{equation} \label{eq:mr}
        mr(c) \;=\; 1 - \frac{\sum\limits_{i = 1}^{c} freq(i)}{N} \;=\;
             \frac{\sum\limits_{i > c} freq(i)}{N}
\end{equation}

The middle form of Equation \ref{eq:mr} counts hits, every reference with stack distance at
most $c$, and subtracts the hit ratio from one. The right-hand form counts the misses
directly: any request for an item with stack distance larger than the cache size results in a
miss, as the item would already have been evicted. First references have infinite stack
distance, so they miss in every cache size.

Some authors write Equation \ref{eq:mr} as an integral of the stack distance density, to
stress that the miss ratio curve is one order (one integration) above the stack distance
distribution. Ding and Xiang's Higher Order Theory of Locality (HOTL)
\cite{xiang2013hotl,ding2012higher} extends this ladder: the footprint is one order above the
miss ratio (Section \ref{sec:hist}).

Table \ref{tab:sdcost} summarizes the cost of selected stack distance algorithms
that we describe in this section.

\begin{table}[t]
    \small
    \centering
    \caption{Cost of selected stack distance algorithms for $N$ references to $M$ distinct data.
    The hash table that stores each datum's last access time ($O(M)$) is
    needed by all tree methods and is not counted in the tree space; SHARDS's own
    table is bounded by $s_{max}$.}
    \label{tab:sdcost}
    \begin{tabular}{@{}lll@{}}
        \toprule
        \textbf{Method} & \textbf{Time} & \textbf{Space} \\
        \midrule
        Mattson \cite{mattson1970evaluation}           & $O(NM)$ & $O(M)$ \\
        Bennett \& Kruskal \cite{bennett1975lru}       & $O(N\log N)$ & $O(N)$ \\
        Olken \cite{olken1981efficient}                & $O(N\log M)$ & $O(M)$ \\
        Kim et al.\ ($S$ sizes) \cite{Kim}             & $O(NS)$ & $O(M)$ \\
        Scale tree, abs.\ error $b$ \cite{ding2003predicting} & $O(N\log (M/b))$ & $O(M/b)$ \\
        Scale tree, rel.\ error $e$ \cite{ding2003predicting} & $O(N\log\log M)$ & $O(\log M)$ \\
        SHARDS, fixed size \cite{waldspurger2015efficient} & $O(N)$ hashing$^{*}$ & $O(1)$ \\
        Increment-and-Freeze \cite{bender2023iaf}      & $O(N\log N)$ & $O(N)$ \\
        \bottomrule
    \end{tabular}

    \vspace{2pt}
    {\footnotesize $^{*}$ plus $O(\log s_{max})$ per \emph{sampled} reference, with $s_{max}$ tracked locations.}
\end{table}

\subsection{Mattson's Stack Distance Algorithm}

Mattson et al.\ \cite{mattson1970evaluation} developed the first algorithm to use
stack distance as a method for calculating the miss ratio for a given cache size. It
has five main steps for each reference $x$:
\begin{enumerate}
    \item Search the stack to find the current location of $x$ (if any)
    \item Calculate the distance $d$ from the top of the stack
    \item Update cache miss counters for all sizes $< d$
    \item Update cache hit counters for all sizes $\ge d$
    \item Remove $x$ from its old location (if any) and push it onto the top of the stack
\end{enumerate}
The algorithm runs in $O(N M)$ time, where $N$ is the total number of references
and $M$ is the number of distinct data. Many works improve on this algorithm
by using more efficient data structures, sampling, and estimated stack distances.

\subsection{Exact Trees}

In tree-based methods we store the LRU stack in a balanced tree, so that finding a datum's
position takes logarithmic rather than linear time. Later tree methods also trade exactness
for space by letting one node stand for many data (Section \ref{sec:approx}).

\paragraph{\textbf{Partial Sums Tree} by Bennett and Kruskal \cite{bennett1975lru}}
Bennett and Kruskal introduce a partial sums tree based on binary vectors representing
the stack distance of a reference. First, they introduce the concept
of using a hash table to store the \emph{previous} location $p$ of
a reference. Second, they maintain a binary vector, $B$, over time,
where $B_i = 1$ if the reference at time $i$ is the \emph{most recent} access
to its datum so far, and $B_i = 0$ otherwise. When $x$ is referenced at time $t$, the
number of 1s in $B$ between $p$ and $t$ is the stack distance of $x$. We then set $B_p = 0$,
since $p$ is no longer the most recent access to $x$, and $B_t = 1$.
To avoid scanning $B$ for each reference, we group $B$ into blocks of 3 and store the number of
1s in each block, group those sums into blocks of 3, and so on. Counting the 1s between $p$ and
$t$ then needs only $O(\log N)$ block sums, and changing one bit updates one sum per level.
For a trace of $N$ references this runs in $O(N \log N)$ time, and since the vector has one
entry per reference, it needs $O(N)$ space.

\paragraph{\textbf{Interval Tree} by Almási et al.\ \cite{Almasi}}
Almási et al.\ interpret Bennett's binary vectors in a different manner. They
record only the 0s from the vector, calling them holes. The calculation for
stack distance now becomes:
\[
    stackdist(x) = t - p - holes(p, t)
\]
where $t$ is the current time, $p$ is the previous time, and $holes(p,t)$ represents
the number of holes (0s) between $p$ and $t$ in $B$. Consecutive holes merge into
intervals, and since the intervals are separated by the last accesses of the $M$ distinct
data, there are at most $M+1$ of them. Stored in an AVL or red-black tree, the intervals take
$O(M)$ space and each reference $O(\log M)$ time, the same bound as Olken's tree below. In
Almási's experiments, recording holes gave a constant factor improvement, and the hole-based
algorithm ran faster than Olken's search tree \cite{zhong2009reusedist}.

\paragraph{\textbf{Search Tree} by Olken \cite{olken1981efficient}}
Olken proposed ordering the LRU stack as a binary tree keyed by
last access time, so that an in-order traversal gives the LRU stack. For
a given node, the right side of the tree represents data
used more recently and the left side represents older data. At each
node in the tree, we store the number of nodes in its subtree.

In order to calculate a stack distance on a reference to $x$, we find the node of its
previous access and start with the size of that node's right subtree. On the way up to the
root, each time we arrive at a parent from its left child, we add one for the parent plus the
size of its right subtree, since all of those data were used more recently. Finally, we add
one for $x$ itself. To keep the tree balanced, Olken used an AVL tree; later implementations
such as PARDA use a splay tree \cite{parda}.
The algorithm runs in $O(N \log M)$ time and needs to store $M$ nodes,
where $M$ is the number of distinct data.

\subsection{Approximate Trees}
\label{sec:approx}

\paragraph{\textbf{Scale Tree} by Ding and Zhong \cite{ding2003predicting,zhong2009reusedist}}
Ding and Zhong observed that we rarely need to know a long stack distance exactly:
whether a reuse has distance one million or 1.01 million does not change our
picture of a program's locality. Their approximate reuse distance analysis, first published in
PLDI 2003 \cite{ding2003predicting} and extended in TOPLAS 2009 \cite{zhong2009reusedist}, where it
is named the \emph{scale tree}, modifies Olken's search tree so that each node represents
a \emph{range} of time instead of the last access of a single datum.
Each node stores the number of data whose last access falls in its time range (its \emph{size}),
the total size of its subtree, and a \emph{capacity}, the largest size the node may reach.

The capacity is set to $d\,e/(1-e)+1$, where $d$ is the number of distinct data accessed
after the node's time range (the stack distance at the node's boundary) and $0<e<1$ is the
error bound chosen by the user. To see why, suppose a node holds $k$ last accesses and our
datum is one of them. We do not know which one, so we assume it is the most recent, and we
undercount its distance by at most $k-1$. In that worst case its true distance is at least
$d+k-1$, so the relative error is at most $(k-1)/(d+k-1)$, and choosing $k = d\,e/(1-e)+1$
makes this exactly $e$. For example, with $e = 1\%$, a node followed by 9,900 distinct data may
hold up to 101 data, while a node near the top of the stack holds just one. Recent time ranges
are therefore kept exact, and older ones are merged into nodes that grow geometrically.

To measure a stack distance, we find the node that holds the previous access and sum subtree
sizes on the way to the root, as in Olken's algorithm. The measured distance is never larger
than the true distance and never smaller than $(1-e)$ times it. For the miss ratio curve, this
means the measured miss ratio at size $c$ lies between the true miss ratios at $c$ and at
$c/(1-e)$. Whenever the tree grows beyond $4\log_{1/(1-e)}M+4$ nodes, a compression pass
merges adjacent nodes whose combined size fits within the capacity. This threshold keeps the
tree at $O(\log M)$ nodes for a fixed $e$ (about 8,250 nodes for $e = 1\%$ and $M = 10^9$), so
each search costs $O(\log\log M)$. Each compression pass at least halves the tree, so its cost
spreads out to $O(1)$ per access, and the total time is $O(N\log\log M)$. Compare this with
$O(\log M)$ per access and $O(M)$ tree nodes for Olken's tree. A second variant bounds the
\emph{absolute} error by a constant $b$, using $O(M/b)$ nodes and $O(N\log(M/b))$ time.

Like every tree method, the scale tree still needs a hash table that records the last
access time of each datum, which takes $O(M)$ space. Ding and Zhong treat this table
separately, and point out that the multi-pass hashing of Bennett and Kruskal can reduce it
\cite{bennett1975lru}. In their experiments, the 99\%-accurate analyzer ran at an almost
constant 1.2 million references per second for data sizes from $10^5$ to $10^{12}$
(with last access times precomputed), and in one comparison used 823 tree nodes where the
exact analyzers needed 67 thousand. This made it practical to analyze whole program
executions, which Ding and Zhong used to predict a program's locality across all of its inputs.
Following this work, Fang et al.\ \cite{fang2005instruction} predicted the stack distance
distribution for the \emph{reference} input of SPEC2K benchmarks using only the
\emph{train} and \emph{test} inputs.
The TOPLAS version also proves that any exact method needs $\Omega(D\log D)$ bits,
where $D$ is the largest stack distance, so Olken's tree is close to optimal for exact
measurement. To our knowledge, the scale tree remains the only approximate method in this
survey with a \emph{deterministic} worst-case error bound; the sampling methods that follow
bound their error statistically.

\subsection{Sampling}

\paragraph{\textbf{SHARDS} by Waldspurger et al.\ \cite{waldspurger2015efficient}}
Spatially hashed approximate reuse distance sampling, or SHARDS, achieves constant space by
sampling references based on the hashed value of their location, hence spatially hashed.
A reference to location $L$ is sampled if $hash(L) \bmod P < T$, giving a sampling rate
$R = T/P$. Because the decision depends only on the location, all references to a sampled
location are kept, so the sampled trace looks like a smaller copy of the original. If a reuse
has true stack distance $d$, about $R\,d$ of the distinct locations accessed in between are also
sampled. So we compute each sampled reuse's distance among the samples with a standard search
tree, and scale it by $1/R$ before adding it to the stack distance histogram.

There are two versions of SHARDS. In the \emph{fixed-rate} version $R$ is constant, so space
grows as $O(RM)$. In the \emph{fixed-size} version, whenever the number of tracked locations
exceeds a bound $s_{max}$, we lower the threshold $T$ to the largest hash value being tracked and
drop that location. The locations that remain are exactly those whose hash is below the new $T$,
so they are still a valid sample, now at a lower rate. Using a fixed size hash table and a fixed
size tree, the fixed-size version runs in constant, $O(1)$, space and requires only one scan
through the trace. The authors also propose SHARDS$_{adj}$, which adjusts the first histogram
bucket by the difference between the expected and the actual number of samples.

\paragraph{\textbf{Improving SHARDS: Skew, Object Sizes and TTLs}}
Spatial sampling is now the default way to build LRU miss ratio curves, and several works
studied where it fails. Carra and Neglia \cite{carra2020efficient} showed that the error of
SHARDS grows with the skew (Zipf exponent) of the workload. Their fix computes the head of the
curve exactly for the most recently used items and joins it to a sampled tail, giving up to
10$\times$ smaller error for about 10\% more CPU. They also propose reporting the error per
miss ratio quantile, since the mean absolute error hides large errors at small cache sizes.
FLOWS by Guo et al.\ \cite{guo2024flows} generalizes this idea: a small exact LRU \emph{filter}
produces the front of the curve and passes only its misses to a sampler whose per-object rate
is proportional to object size. For byte miss ratio curves at a 1\% sampling rate, FLOWS lowers
the error from 7.29\% (Carra and Neglia's method) to 0.44\%.
Finally, Sultan et al.\ \cite{sultan2024ttls} showed that time-to-live (TTL) expiration changes
the curve itself. On 28 Twitter workloads, sizing with TTLs needed 69\% less memory on
average (up to 99\%) for the same hit ratio. They make Mattson's, Olken's and SHARDS'
algorithms TTL-aware by adding an expiration-time priority queue.

\paragraph{\textbf{Sampling for Programs}}
Several earlier works also reduce the cost of program reuse distance analysis through
sampling. Zhong and Chang \cite{zhong2008sampling} sample the trace and report over 99\%
accuracy with a 7.5$\times$ average speedup. Schuff et al.\ \cite{schuff2010multicore}
combine sampling with parallelization for multicore caches.

\subsection{Parallel Analysis}

\paragraph{\textbf{PARDA} by Niu et al.\ \cite{parda}}
Q.\ Niu et al.\ introduce a parallel stack distance algorithm that partitions
the trace into chunks, one per processor. Each processor computes the stack distances
in its chunk with its own hash table and splay tree. References whose previous use lies in an
earlier chunk appear as local cold misses, and are resolved by passing them, in order, to the
processor that holds the earlier chunk, which replays them on its own stack.
Cui et al.\ \cite{cui2012hprda} compute reuse distances on GPUs, 20$\times$ faster with less than
1\% loss in precision.

\paragraph{\textbf{Increment-and-Freeze} by Bender et al.\ \cite{bender2023iaf}}
Increment-and-Freeze computes \emph{exact} stack distances by divide and conquer over an array
of per-request counters, with range increment and freeze operations, instead of a tree. It runs
in $O(N\log N)$ time, is I/O efficient, and parallelizes well: it is 4--9$\times$ faster than a
tree on one core, up to 60$\times$ faster on 16 threads, and 2--10$\times$ faster than PARDA.
As Bender et al.\ note, achieving a small \emph{multiplicative} error requires linear space,
which limits what any sketch can promise; note that the scale tree's tree is sublinear, but its
table of last access times is not.

\subsection{Buckets}

These algorithms group the LRU stack into a small number of buckets while still
maintaining the LRU stack order between buckets.

\paragraph{\textbf{MIMIR} by Saemundsson et al.\ \cite{saemundsson2014cloudcaches}}
The MIMIR algorithm creates a fixed number of buckets to store references. Each bucket
$B_i$ maintains the LRU stack order property, that is $B_i$ has lower stack distance
than $B_j$ if $i < j$. While buckets can vary in size, they introduce an
\emph{aging} procedure to ensure that buckets stay balanced.
When a request hits a datum in bucket $B_i$, we only know its stack distance lies between
$n_i+1$ and $n_i+|B_i|$, where $n_i$ is the total size of the buckets above $B_i$. MIMIR spreads
that hit evenly over the range, and the miss ratio for a cache size $c$ is then computed from
the resulting stack distance histogram with Equation \ref{eq:mr}.

\paragraph{\textbf{Cliffhanger} by Cidon et al.\ \cite{cidon2016cliffhanger}}
Cliffhanger does not build a full curve. Instead, it uses \emph{shadow queues}, which can be
thought of as victim caches that store only the \emph{key} of the data, not the value. A hit in
a shadow queue is a request that missed but would have hit if the cache were a little larger,
so the rate of shadow-queue hits estimates the slope of the hit ratio curve at the current
size. Cliffhanger uses \emph{hill climbing}: it moves memory toward the cache with the steepest
slope, taking it from a randomly chosen other cache, until no move improves the total hit ratio.
Memshare \cite{cidon2017memshare} applies the same idea to multi-tenant key-value caches.

\subsection{Histogram}

These methods either use: (1) a set of counters to represent stack distances at
each cache size or (2) a set of counters to represent \emph{total} accesses
and \emph{unique} accesses. The counts then represent a histogram of the stack distances
that is used to construct the miss ratio curve.

\paragraph{\textbf{Powers of Two} by Kim et al.\ \cite{Kim}}
This method by Kim et al.\ changes Mattson's stack processing algorithm by
recording stack distances only at a set of $S$ cache sizes, for example powers of 2.
The idea is based on the insight that memory designers often do not require miss
ratios for continuous memory sizes; a few sizes are sufficient.
The algorithm stores the LRU stack as a linked list with a marker at each of the $S$ chosen
sizes, and a hash table is used to find the location of references in the stack.
On a reference, only the markers above the referenced datum move, so the algorithm runs
in $O(NS)$ time, linear in the trace length and independent of the trace's locality.
Note that it gives exact miss counts at the $S$ chosen sizes only, not the full
stack distance distribution.

This algorithm has been applied to build miss ratio curves in Zhou et
al.\ \cite{zhou2004dynamic}. They present
a hardware and software implementation for calculating page miss ratio
curves to guide memory management in a multi-program environment.
At the hardware level, Tam et al.\ \cite{Tam} have used this algorithm
to manage L2 cache partitioning. Here they create
an access trace to L2 by recording the misses of L1, and then they feed the access trace
into the algorithm given by Kim. Their miss ratio
curve is then used to guide how much of the L2 cache an application should
receive.

\paragraph{\textbf{UMONs} by Qureshi and Patt \cite{qureshi2006utility}}
Qureshi and Patt developed utility monitor circuits (UMONs) that
approximate hit ratio vs.\ number of ways in an LRU cache, and perform way-partitioning
based on the approximated hit ratio curves.
UMONs use a counter at each stack position (way) of a set of shadow tags.
Since the cache obeys the LRU policy, the hit ratio for a $w$-way partition is found by
summing the counters up to $w$ ways. If application A has a higher utility for a
number of ways (i.e.\ its hit ratio curve is steeper) then it will be assigned
those ways. Application B will be left with the remaining number of ways.

\paragraph{\textbf{Counter Stacks} by Wires et al.\ \cite{wires2014characterizing}}
Counter Stacks keeps a stack of counters. A new counter starts at regular intervals, and each
counter records how many distinct data have been accessed since it started. The counters are
probabilistic HyperLogLog sketches, so each takes only a few kilobytes no matter how many data
it counts. Table \ref{tab:cs} shows the counters for the requests a, b, c, a, with a new
counter started at every request.

\begin{table}[t]
\small
\centering
\caption{Example Counter Stacks calculation. Row $i$ is the counter started at time $i$;
each column gives the counter values after the request at the top.}
\label{tab:cs}
\begin{tabular}{llll}
\toprule
    \textbf{a} & \textbf{b} & \textbf{c} & \textbf{a} \\
\midrule
1 & 2 & 3 & 3 \\
  & 1 & 2 & 3 \\
  &   & 1 & 2 \\
  &   &   & 1 \\
\bottomrule
\end{tabular}
\end{table}

To find a request's stack distance, we look at which counters grew. The counter started at
time $i$ grows on a request only if the requested datum has not been seen since time $i$. When
$a$ is requested at time 4, the counter started at time 1 stays at 3, because $a$ was already
counted, but the counter started at time 2 grows from 2 to 3. So the previous access to $a$
was at time 1, and its stack distance is the value of the time-1 counter, 3 ($a$, $b$ and $c$).
In practice a new counter starts only every few thousand requests, so we learn how many requests
had distances in each range between two counters rather than each request's exact distance. A
counter is dropped once its value comes close to that of the next older counter, which keeps
the number of counters, and so the space, sublinear.
Sultan et al.\ \cite{sultan2024ttls} extend HyperLogLog counters with expiration times
(CounterStacks++), which reach 0.32--0.39\% average error with 50 counters on TTL workloads.
Drudi et al.\ \cite{drudi2015approximating} give the first provable sublinear-space
approximation of LRU hit ratio curves in the streaming model, and Arima and Schulz
\cite{arima2025} relate miss curves to cardinality curves that can be fitted with the same kind
of counting structures.

\paragraph{\textbf{Fractals} by He et al.\ \cite{fractal}}
L.\ He et al.\ compute the miss ratio from the time between two consecutive misses, known as
the inter-miss gap described by Denning and related to LRU miss ratio by Ding in HOTL
\cite{xiang2013hotl}. The average inter-miss gap of a cache of size $c$ is $1/mr(c)$, and they
use a fractal model of how the inter-miss gap distribution changes with cache size to predict
the miss ratio of other cache sizes.
FractalMRC is implemented using hardware performance counters, and their model allows for
online processing at the cost of a 2\% slowdown.
One drawback is that their model ignores cold misses, which leads to inaccuracy (the cold miss
equation is not fractal). On average in the SPEC2006 benchmark they report 76\% miss ratio
curve accuracy.

\subsection{Symbolic Analysis}

The largest new branch since our first version is \emph{static} analysis of
affine loop programs, which computes stack distances or miss counts without running a trace.
PolyCache \cite{bao2018polycache} gives closed-form miss counts for set-associative
cache hierarchies. HayStack \cite{gysi2019haystack} uses Barvinok counting twice: once to derive
each access's stack distance as a symbolic expression, and once to count the accesses whose
distance exceeds the cache size. It often finishes within seconds, mostly independent of the
problem size, with less than 0.6\% error on real hardware, but it models only fully associative
LRU caches. Warping \cite{morelli2022warping} instead fast-forwards an explicit cache simulation
of polyhedral programs, so it supports real replacement policies.
These methods trade generality (affine loops only) for a cost that does not depend on the
trace length.

\section{Reuse Time Algorithms}

More recently, reuse time has emerged as a metric used to approximate miss ratio
curves. Denning and Schwartz \cite{denning1972properties} showed in 1972 that the reuse time
distribution yields the average working set size, but until recently it was hard to achieve
accurate miss ratio curves with the analytical models available.
Table \ref{tab:rt} gives an example of the reuse time calculation for the requests of
Table \ref{tab:sd}. Compare the last rows of the two tables: $b$ has reuse time 6 but stack
distance 4, since two of the requests in between were repeats. We do not need to maintain
the LRU stack, but as a result the miss ratio for a given cache size is no longer an exact sum
over the distribution of reuse times.

\begin{table}[t]
    \small
    \centering
    \caption{Example reuse time calculation for the requests of Table \ref{tab:sd}.}
    \begin{tabular}{llll}
    \toprule
    \textbf{request} & \textbf{time} & \textbf{last use} & \textbf{reuse time} \\ \midrule
    a & 1 & -- & $\infty$ \\
    b & 2 & -- & $\infty$ \\
    c & 3 & -- & $\infty$ \\
    d & 4 & -- & $\infty$ \\
    a & 5 & 1  & 4 (5$-$1) \\
    a & 6 & 5  & 1 (6$-$5) \\
    d & 7 & 4  & 3 (7$-$4) \\
    b & 8 & 2  & 6 (8$-$2) \\
    \bottomrule
    \end{tabular}
    \label{tab:rt}
\end{table}

\subsection{Histogram}
\label{sec:hist}

These methods bin the data recorded into a histogram, and build
a miss ratio curve from that histogram by integrating the
probability density function or summing the total frequencies
over a given range.

\paragraph{\textbf{Average Eviction Time} by Hu et al.\ \cite{hu2016kinetic,hu2018fast}}
Hu et al.\ follow a datum as it moves down the LRU stack after its last access. At each later
access, our datum moves down one position if the requested datum was below it in the stack,
or is new. That happens exactly when the requested datum's reuse time is longer than the time
since our datum's last access. So, $t$ accesses after its last use, our datum moves down with
probability $P(rt > t)$, which we read from the reuse time histogram. After $T$ accesses it has
moved down $\int_0^{T}P(rt > t)\,dt$ positions, and it falls out of a cache of size $c$ when this
sum reaches $c$. That time is the \emph{average eviction time}, $AET(c)$:
\begin{equation}
    \label{eq:aet}
    \int_0^{AET(c)} P(rt > t)\,dt = c
\end{equation}
A reference misses if its datum was evicted before it came back, that is, if its reuse time
is longer than $AET(c)$:
\begin{equation}
    \label{eq:aetmr}
    mr(c) = P(rt > AET(c))
\end{equation}

To control space and time overhead, AET adopts random sampling and reservoir sampling to
generate the reuse time histogram, and maintains a small hash table of the sampled keys to
compute their reuse times. AET can generate an accurate MRC with a sampling ratio as low
as 1/10000. The journal version of AET \cite{hu2018fast} adds adaptive phase sampling and
models for shared and multi-level exclusive caches. On the 2.4-billion-access MSR trace,
with 8K-entry tables for both, AET reached a mean absolute error of 0.01 in 1.7~MB at
79 million blocks per second, compared with 0.006 in 2.3~MB at 81 million blocks per second for
SHARDS. In other words the two families have converged in cost; AET distinguishes itself
by composition and phase handling, and SHARDS by its generality.

\paragraph{\textbf{Footprint} by Xiang et al.\ \cite{xiangWSS}}
Xiang et al.\ introduce the \emph{footprint} function as a mapping from an
execution window of length $x$ to the expected amount of data accessed during the window,
$fp(x)$. A trace of $N$ references has about $N^2/2$ windows, so measuring each one would take
quadratic time. Xiang et al.\ show that we can compute $fp(x)$ for every window length in
linear time by counting, for each datum, the windows that do \emph{not} touch it. A window of
length $x$ misses a datum only if it fits inside one of the datum's reuse intervals, before its
first access, or after its last access. So we only need to measure:
\begin{itemize}
    \item Reuse Time Histogram
    \item The first use of every datum
    \item The last use of every datum
\end{itemize}
The miss ratio for cache size $c$ is the finite difference of the footprint
at the window length where the footprint equals the cache size \cite{xiang2013hotl}:
\begin{equation}
    \label{eq:hotl}
    mr(c) = fp(x+1) - fp(x) \quad \text{where } fp(x) = c
\end{equation}
The difference $fp(x+1)-fp(x)$ is the chance that one more access touches new data. When the
cache is just large enough to hold the footprint of $x$ accesses, touching new data is a miss.
In HOTL, it was shown that the footprint metric is the cache locality theory equivalent of
Denning's working set size \cite{xiang2013hotl}.
The footprint algorithm runs in $O(N)$ time and $O(M)$ space. The
algorithm was applied in LAMA (Locality Aware Memory Allocation) \cite{hu2015lama,HU:TC17}
to guide \verb!memcached!'s slab reallocation problem.

\paragraph{\textbf{Hit-Evict Distribution} by Beckmann and Sanchez \cite{beckmann2016modeling}}
Beckmann and Sanchez describe a cache with three distributions over \emph{age}, the time since
a line was last referenced: the ages of lines in the cache, the ages at which lines hit, and the
ages at which they are evicted. The replacement policy enters as a ranking over ages. Solving
the three together from the reuse time distribution gives the miss ratio of any age-based
policy, not only LRU. The motivation behind this work is to reduce the dependence on the LRU
stack property that many analytic models assume.

\subsection{Access Counters}

These methods use counters to record reuse times for a particular
set of references; these counters are then aggregated to form a reuse
time distribution.

\paragraph{\textbf{StatCache} by Berg and Hagersten \cite{berg2004statcache}}
StatCache uses random sampling to select some memory references
for its analysis. For each memory reference sampled, it does the following:
\begin{enumerate}
    \item Set a (hardware) watchpoint on the address and record the current number
          of memory accesses at that time, $t_0$.
    \item On the watchpoint trap read the new current number of memory accesses, $t_1$.
          The reuse time of this address is now $t_1-t_0$.
    \item Update the reuse time distribution with the new reuse time.
\end{enumerate}
The StatCache model assumes a cache of $c$ lines with random replacement, where each miss
evicts a line chosen uniformly at random. Suppose the miss ratio is $mr$. A reuse with reuse
time $t$ sees about $t\cdot mr$ misses in between, and our line survives each of them with
probability $1-1/c$, so the reuse misses with probability $1-(1-1/c)^{t\cdot mr}$. Averaging
this over the reuse time histogram must give back the miss ratio we started with:
\[ mr = \sum_t P(rt = t)\,\bigl[1-(1-1/c)^{t\cdot mr}\bigr], \]
where first accesses ($t=\infty$) always miss. StatCache solves this equation for $mr$
numerically, once for each cache size.

\paragraph{\textbf{StatStack} by Eklöv and Hagersten \cite{eklov2010statstack}}
The StatStack model builds on StatCache, this time for an LRU cache. Its key step is to
estimate stack distances from reuse times alone. Take a reuse with reuse time $t$. An access
$i$ steps before the end of the window adds a new distinct datum to the window only if that
datum is not accessed again before the window ends, which happens with probability
$P(rt > i)$. Adding these up, the expected stack distance of a reuse with reuse time $t$ is
about $\sum_{i<t}P(rt>i)$. StatStack converts each sampled reuse time into an expected stack
distance this way, which gives a stack distance histogram, and then applies
Equation \ref{eq:mr}.

\paragraph{\textbf{Hardware Sampling}}
RDX by Wang et al.\ \cite{wang2019rdx} combines performance monitoring unit (PMU) sampling with
hardware debug registers, which play the role of StatCache's watchpoints, and needs no
instrumentation. It costs about 5\% time and 7\% memory overhead with over 90\% accuracy.
ReuseTracker \cite{sasongko2021reusetracker} extends the approach to multithreaded programs.
Although RDX reports reuse distances, it samples reuse \emph{times} and converts them, so we
place it with StatStack.

\paragraph{\textbf{Re-access Ratio} in Block Storage}
Cloud block storage vendors built inline estimators that keep a little state for every block
instead of sampling. OSCA \cite{zhang2020osca} keeps two counters per block, fits the
re-access ratio as a function of reuse time, and from the fit estimates each reuse distance in
$O(1)$. On 20 production nodes it reached a mean absolute error of 0.005, compared with 0.017
for miniature simulation, and its cache reallocation cut back-end traffic by 13.2\%.
APAC \cite{apac2022} estimates reuse distances probabilistically in linear time and reports
lower error than both SHARDS and OSCA on block storage traces. Cuki \cite{gu2023cuki} tracks
the working set size and the item repetition ratio over a sliding window using a
cuckoo-filter-like table, so the curve reflects only recent requests.

\subsection{Static Sampling}

Static Parallel Sampling (SPS) by Chen et al.\ \cite{chen2018sps} has the compiler generate
sampler code that computes each sampled reuse time in $O(1)$ from the loop structure. Sampling
2\% of loop iterations gives almost exact miss ratio curves, at a small fraction of the cost of
tracing. PLUSS \cite{liu2024pluss} extends this to parallel OpenMP loops with a symbolic thread
count, reaching about 97\% MRC accuracy and re-deriving the curve for a new thread count in
a fraction of a second.

\subsection{Toward a Unified Model: Reuse Time, Footprint and Eviction Time}
\label{sec:unify}

Since our first version, there has been work on unifying cache modeling techniques. In the first version
we noted that if one were to interpret StatStack's relationship between average stack distance
and time, one would arrive at the kinetic equation of AET. Yuan, Ding, Denning and Zhang made
this precise in the Measurement Theory \cite{yuan2018measurement} and Relational Theory
\cite{yuan2019relational} of locality. They prove that the working set recursion of Denning and
Schwartz \cite{denning1972properties},
\[
   s(x) = \sum_{i<x} P(rt > i),
\]
equals the steady-state footprint for every trace, with no stochastic assumption.

To see that these are one model, put this sum next to Equations \ref{eq:aet}--\ref{eq:hotl}.
The sum $s(x)$ is the discrete form of the integral in Equation \ref{eq:aet}, so $AET(c)$ is
the window length at which $s$ reaches $c$: AET is the fill time, the inverse of the
footprint. Equation \ref{eq:aetmr} says $mr(c) = P(rt > AET(c))$, and
$P(rt > x) = s(x+1)-s(x)$ is exactly the finite difference in Equation \ref{eq:hotl}.
StatStack's expected stack distance for a reuse time $t$ is, up to its first term, the same
sum $s(t)$. Shen et al.'s locality approximation \cite{shen2007locality}, which converts
reuse times into reuse distances and was 17$\times$ faster than measuring distances directly,
is the same recursion normalized differently. StatStack, footprint differentiation and AET
integration are therefore \emph{three algorithms for one model}, which differ mainly in how they
treat the first and last accesses and cold misses.

Measurement Theory also shows where the conversion can fail. The reuse time histogram of an interleaved
trace is the sum of the rate-normalized histograms of its components, but the stack distance
histogram is not composable, even when the interleaving is known. This is why the shared cache
models since 2017 work in reuse time. Brock et al.\ \cite{brock2018composable} proved the
footprint composition for uniform interleaving with a tight error bound, and Hu et al.\
\cite{hu2017symbiosis} used footprint composition to pick co-run groups within 1.5\% of the best.
The footprint was also extended to exclusive and victim caches \cite{ye2017vfp}, to threads that
share data \cite{luo2017shared}, and to writebacks \cite{smith2021writeback} (Section \ref{sec:wbr}).

The theory also found the \emph{conditions} under which these models are exact. Chen et al.\
\cite{chen2021ismm} proved that, under their Maximal RI Sum assumption, HOTL computes exactly
the miss ratio of a cache with uniform leases, which explains its two known failure modes:
over-prediction on sawtooth access patterns and under-prediction on phased ones. For AET, the
model is exact at cache size $c$ if and only if two kinds of error balance: reuses predicted as
misses (reuse time above $AET(c)$ but stack distance within $c$), and reuses predicted as hits
\cite{hu2018fast}. We note that no unconditional bound on the error of HOTL or AET against LRU
is known.

\section{Beyond LRU}
\label{sec:nonlru}

In our first version, the only non-LRU entries were StatCache (random replacement) and
the hit-evict model of Beckmann and Sanchez, and we listed non-LRU caches as an
open problem. Since then, curves for arbitrary policies have been built mainly by
scaled-down simulation (Figure \ref{fig:any}).

\paragraph{\textbf{Miniature Simulation} by Waldspurger et al.\ \cite{waldspurger2017minisim}}
Miniature simulation (MiniSim) turns the SHARDS idea of \emph{sample, then compute stack
distances} into \emph{sample, then simulate a shrunken cache}. A cache of size $c$ is emulated by
a mini-cache of size $Rc$ fed the spatially sampled references at rate $R$, using any
replacement policy. Misses are divided by the expected rather than the actual number of
sampled references to cancel sampling bias. Across 137 traces, $R=0.001$ gave a median mean
absolute error below 0.005 for ARC, LIRS and OPT, using about 1000$\times$ less space than full
simulation. Since each point of the curve needs its own mini-cache, it is also the cost baseline
that later work tries to beat. The same paper documents non-monotonic miss ratio curves for
LIRS and ARC, and proposes SLIDE, which removes performance cliffs in the style of Talus
\cite{beckmann2015talus}. The libCacheSim simulator \cite{libcachesim} includes a miniature
simulation profiler for its many policies, which is how many researchers obtain non-LRU curves
today.

\paragraph{\textbf{Kosmo} by Shakiba et al.\ \cite{shakiba2024kosmo}}
Kosmo gives each object an \emph{eviction map} recording when, and in what state, it would
have been evicted from caches of each size. From a single copy of the metadata, Kosmo
reconstructs the cache contents of LFU, FIFO, 2Q, LRFU, LRU and MRU caches of many sizes in one
pass. Against miniature simulation at 100 cache sizes, it used 3.6$\times$ less memory on
average with mean absolute error within 0.25\%. Kosmo also measured how often each policy
violates the inclusion property between cache sizes: 1.5\% of accesses for LFU, but 20.6\% for
FIFO and 29.1\% for MRU. Its curves are monotone by construction, which misrepresents
non-monotonic policies, and policies with multiple stacks such as ARC are left for future work.

\paragraph{\textbf{Other Approaches}}
KRR \cite{yang2021krr} generalizes Mattson's stack to the sampled LRU of Redis, which evicts the
least recent of $K$ random samples; with $K=1$ it models random replacement. On the analytical
side, the characteristic time (TTL) approximation of Che et al.\ \cite{che2002hierarchical}
models an LRU cache as one that keeps each object for a fixed time after its last request. It
has been extended to multi-list LRU variants \cite{gast2017lrum}, proved asymptotically exact for
LRU under independent references \cite{jiang2018ttl}, and very recently applied to the CLOCK
family \cite{wang2026clockk}; a working-set approximation has also been proposed for SIEVE
\cite{tanvir2026sieve}. These models assume
independent references. Sen and Wood \cite{sen2013} model set-associative caches with several
hardware policies from reuse information. Finally, LPCA \cite{gu2022lpca} claims to be the
first work to learn miss ratio curves for non-LRU policies with machine learning.

\paragraph{\textbf{Optimal Replacement}}
The optimal policy, MIN or OPT, evicts the object whose next use is farthest in the future
\cite{belady1966study}. For objects of equal size it has the fewest misses at every cache size,
so its curve is a lower bound on the curve of every replacement policy, and the gap between the
LRU and OPT curves shows how much a better policy could gain. Mattson et al.\
\cite{mattson1970evaluation} showed that OPT is also a stack algorithm: if each object's
priority is its next use, one stack pass gives the whole curve, once the trace has been scanned
ahead to find those next uses. Sugumar and Abraham \cite{sugumar1993opt} simulated many OPT
caches in one pass with a limited lookahead, and found OPT miss ratios up to 70\% lower than
LRU in fully associative caches. Although OPT needs the future to decide what to evict, the
stack distance of an access depends only on the accesses before it, so it can be computed
online. Bilardi et al.\ \cite{bilardi2017csa} do so in $O(\log M)$ time per access, and prove a
matching lower bound that holds even for simulating a single cache size offline. Hawkeye
\cite{jain2016hawkeye} relies on the same fact in hardware: it computes OPT's decisions for
past accesses and trains its replacement policy to imitate them. When object sizes vary,
computing OPT is NP-hard. FOO \cite{berger2018foo} casts caching as a min-cost flow problem to
bound OPT from above and below, and its scalable variant shows that production caching systems
have 11--43\% more misses than OPT.

\paragraph{\textbf{Multi-tier Caches}}
eMRC by Liu et al.\ \cite{liu2021emrc} builds a multi-dimensional miss ratio
function over the sizes of every tier of a multi-tier cache, where each tier has its own
replacement policy. It is not a new single-tier algorithm; instead, it generalizes the cliff
removal of Talus to several dimensions, and uses a convex hull approximation so that only a few
configurations need to be simulated with SHARDS or miniature simulation. It needs 14$\times$
fewer simulated points for two tiers and 4,527$\times$ fewer for four tiers than exhaustive
simulation, with less than 2\% mean absolute error. Estro et al.\ \cite{estro2023} propose using
knee detection for the same problem.

\section{Writeback Ratio Curves}
\label{sec:wbr}

Miss ratio curves count how often data must be \emph{read} into a cache. For memory and storage
devices with limited write endurance or bandwidth, such as non-volatile memory (NVM) and flash,
we also need to know how often data must be \emph{written} out of the cache. In a write-back
cache, a write marks the cached object \emph{dirty}, and the object is written back to the next
level only when it is evicted. Each miss causes an eviction, but only the eviction of a dirty
object causes a writeback. We define the \emph{writeback ratio} $wbr(c)$ as the number of
writebacks over the total number of requests for a cache of size $c$. Since every writeback
is an eviction caused by a miss, $wbr(c) \le mr(c)$, as shown in Figure \ref{fig:wbr}.

\begin{figure}[t]
\centering
\begin{tikzpicture}
\begin{axis}[
    width=\columnwidth, height=5.4cm,
    xmin=0, xmax=100, ymin=0, ymax=0.6,
    xlabel={Cache size (\% of data size)}, ylabel={Ratio of requests},
    grid=major, grid style={gray!20},
    tick label style={font=\footnotesize}, label style={font=\footnotesize},
    legend style={font=\scriptsize, draw=none, fill=white, fill opacity=0.8, text opacity=1},
    legend cell align=left]
  \addplot[very thick, blue!70!black] table[x=pct, y=mr] {figures/data/wbr_sim.dat};
  \addlegendentry{miss ratio $mr(c)$}
  \addplot[very thick, dashed, red!70!black] table[x=pct, y=wbr30] {figures/data/wbr_sim.dat};
  \addlegendentry{writeback ratio, 30\% writes}
  \addplot[only marks, mark=o, mark size=1.6pt, red!70!black] table[x=pct, y=wbr30] {figures/data/wbr_model.dat};
  \addlegendentry{Equation \ref{eq:wbr} model, 30\% writes}
  \addplot[very thick, dotted, orange!80!black] table[x=pct, y=wbr10] {figures/data/wbr_sim.dat};
  \addlegendentry{writeback ratio, 10\% writes}
  \draw[{Stealth}-{Stealth}, black!70] (axis cs:20,0.1268) -- (axis cs:20,0.3291);
  \draw[black!50, thin] (axis cs:20.5,0.235) -- (axis cs:49,0.30);
  \node[font=\scriptsize, anchor=west, align=left, black!80] at (axis cs:49,0.30)
      {misses that evict\\a clean object};
\end{axis}
\end{tikzpicture}
\caption{Miss ratio and writeback ratio of an LRU write-back, write-allocate cache. The
workload is synthetic: 1 million requests over 10,000 objects (Zipf, $\alpha=0.9$); each
request is a write with probability 10\% or 30\%. A miss causes a writeback only when the
object it evicts is dirty, so both writeback curves lie below the miss ratio curve. The arrow
marks, for 30\% writes, the misses that evict a clean object. Circles show the prediction of
Equation \ref{eq:wbr}, which is within 2.2\% of the simulation for cache sizes up to 90\%.}
\Description{Three curves fall as cache size grows: the miss ratio on top, the writeback ratio
with 30\% writes in the middle, matched by model circles, and the writeback ratio with 10\%
writes at the bottom.}
\label{fig:wbr}
\end{figure}

Chen et al.\ \cite{chen2016write} introduce a metric for write locality, the
\emph{write reuse interval}. Take two consecutive writes to an object $a$, $a_1$ and $a_2$, with
any number of reads of $a$ in between. Each access after $a_1$, up to and including $a_2$, has a
reuse time: the gap since the previous access to $a$. The write reuse interval is the longest of
these gaps. If it is longer than the cache's eviction time, $a$ is evicted somewhere in the span
while it is dirty, which costs one writeback. Chen et al.\ construct a write reuse time histogram
and combine it with the footprint to compute the writeback ratio in linear time, and apply it to
partition cache space among co-run programs on hybrid DRAM and NVM main memory.

Smith, Byrne and Ding \cite{smith2021writeback} model writebacks directly from the footprint,
assuming the next level is large enough to hold everything. They define the \emph{write
footprint} $wfp(x)$ as the expected number of distinct objects written in a window of $x$
requests, and the dirty ratio $dfp(x) = wfp(x)/fp(x)$. When the cache holds the footprint of
$x$ requests, each miss evicts one object, and that object is dirty with probability about
$dfp(x)$:
\begin{equation}
    \label{eq:wbr}
    wbr(c) = mr(c) \cdot dfp(x) \quad \text{where } fp(x) = c
\end{equation}
For Zipfian workloads both terms have closed forms, so the model needs no trace at all.
It averaged 5.0\% relative error, and is faster than trace
analysis for a single cache size. A second, more general model counts the accesses that will
eventually incur a writeback, and works for any size of the next level. It averaged 3.0\% relative
error and was at least 105$\times$ faster than using trace-based methods.

\section{Miss Ratio Curves in Practice}

\paragraph{\textbf{Key-Value Caches}}
Miss ratio curves have been used in \emph{memcached} for provisioning slab
classes. In memcached each object belongs to a class based on its size,
and so MRCs can be used to maximize hit ratio over all classes for a given amount of memory.
LAMA \cite{hu2015lama,HU:TC17} achieves this by using the footprint.
MIMIR \cite{saemundsson2014cloudcaches} and Dynacache \cite{cidon2015dynacache} also approach the
problem by modeling the miss ratio for each slab class. Among multiple tenants, mPart
\cite{byrne2018mpart} builds AET curves online for each tenant and partitions memory to maximize
the overall hit ratio, and pRedis \cite{pan2019predis} weights each tenant's curve by its miss penalty.
Other systems use only the local gradient of the curve: Memshare \cite{cidon2017memshare} uses
shadow queues, and the CacheLib \cite{cachelib2020} rebalancer equalizes the marginal hits in the
LRU tail of each slab class.

\paragraph{\textbf{Virtual Machines and Storage}}
In the context of virtualization, memory allocation between virtual machines on a single
host has been studied extensively \cite{zhao2009dynamic}. Wang et al.\ \cite{TACO16} use the
miss ratio curve in a dynamic programming memory balancer that minimizes the expected number of page
misses, which in turn guides virtual machine memory allocation in a cloud environment. In cloud
block storage, OSCA \cite{zhang2020osca} reallocates cache among volumes using online curves.

\paragraph{\textbf{Production Models}}
At large scale, curves are often replaced by parametric models. Akamai's footprint descriptors
predict CDN cache requirements within 2.5\% \cite{sundarrajan2017footprint}, Alibaba's iBTune fits
a power law of miss ratio against buffer size for more than 10,000 database instances
\cite{tan2019ibtune}, and Google's CacheSack models LRU as a TTL cache rather than simulating
thousands of categories \cite{yang2022cachesack}. FaasCache \cite{fuerst2021faascache} uses hit ratio
curves to size the memory that keeps serverless functions warm.

\paragraph{\textbf{Operating Systems and Tiered Memory}}
Operating systems estimate the \emph{cold} fraction of memory rather than full curves.
Google's far memory \cite{lagarcavilla2019far} builds per-job histograms of page age from
page table accessed bits, and picks the age threshold for moving pages to far memory so that
promotions stay below 0.2\% of the working set per minute. This is effectively a
miss ratio curve in physical time. Linux's multi-generation LRU \cite{mglru2026} and DAMON
\cite{park2019damon} expose similar access histograms, while Meta's TMO \cite{weiner2022tmo} uses
memory pressure and builds no curve at all. Tiered memory systems such as HeMem
\cite{raybuck2021hemem} and Memtis \cite{lee2023memtis} size the hot set from page access
\emph{frequency} histograms, and TPP \cite{maruf2023tpp} places pages using LRU-based demotion and
promotion on access.

\paragraph{\textbf{Processor Caches}}
Commodity processors do not expose hardware miss curve monitors such as UMONs.
Instead, with Intel's cache allocation technology, KPart \cite{elsayed2018kpart} measures a few
way allocations with performance counters to build the curve, and FLORIA \cite{floria2023}
obtains a per-set miss distribution by sampling selected cache sets. Jenga \cite{tsai2017jenga}
uses hardware monitors in a simulated design to build software-defined cache hierarchies.

\section{Open Problems}

We revisit two of the open problems from the first version of this survey. Both have seen
progress since, but neither is closed.

\begin{itemize}
    \item \textbf{To what extent can we relate logical access time to physical clock time?} \\
          Composition models bring physical access rates in by stretching time
          \cite{hu2017symbiosis,hu2018fast}, and timescale theory treats a window as
          either physical or logical time \cite{yuan2019relational}. In practice, Google's far
          memory curves are in seconds \cite{lagarcavilla2019far}, TTL-aware stacks put wall-clock
          expiration inside a logical time simulation \cite{sultan2024ttls}, and phases are handled
          by building one histogram per phase \cite{hu2018fast}. What is still missing is a
          validated, general conversion from a logical time MRC to physical miss rates
          (misses per second).
    \item \textbf{How to extend these models to non-LRU caches?} \\
          This is now solved in practice for many policies by scaled-down simulation
          and single-pass reconstruction \cite{waldspurger2017minisim,shakiba2024kosmo}.
          Still open are single-pass curves for adaptive policies such as ARC and LIRS, an error
          theory for scaled-down simulation, and a faithful representation of non-monotonic
          curves. Contrary to our expectation in the
          first version, Beckmann's follow-up work produced replacement policies
          \cite{beckmann2017eva,beckmann2018lhd} rather than policy-specific models.
\end{itemize}

We also present the following areas of future work that would help advance the field:

\begin{itemize}
    \item \textbf{Guarantees.} The techniques in this survey offer very different
          guarantees (Table \ref{tab:glance}): deterministic (the scale tree), statistical
          (SHARDS), randomized with provable bounds (Drudi et al.), asymptotically exact
          under independent references (TTL models), exact under checkable conditions (HOTL,
          AET), or purely empirical (miniature simulation, learned models). Since a small
          multiplicative error requires linear space, is there a \emph{deterministic}
          sublinear-space algorithm with additive error?
    \item \textbf{Bytes, TTLs and tiers.} Byte miss ratio, TTL expiration and multiple tiers each
          change the curve \cite{guo2024flows,sultan2024ttls,liu2021emrc}, but no single method
          handles all three, together with writebacks, for non-LRU policies.
    \item \textbf{Memory-side counters.} CXL memory controllers can now count accesses for every
          page \cite{sun2025m5}. Counts alone give frequency curves, not LRU curves, but combined
          with sampled reuse times they could make page-level miss ratio curves cheap for
          tiered memory.
\end{itemize}

\section{Conclusion}

Our survey has covered the two main metrics, stack distance and reuse
time, and their associated cache models from a miss ratio curve construction
standpoint, as well as the newer techniques that simulate or model the cache state of any
replacement policy. In Denning's 1968 work \cite{denning1968working}, he
defined the working set as the data used in a window of time. The footprint is its average
size, and the average eviction time is the window length whose working set just fills the
cache. The relational theory of locality shows that these, together with StatStack's expected
stack distance, are one model. You could say that modeling locality has been reused over and
over again.

\bibliographystyle{ACM-Reference-Format}
\bibliography{mrc}

\end{document}